%% file: main.tex
\documentclass[conference]{IEEEtran}
\IEEEoverridecommandlockouts
\usepackage{cite}
\usepackage[nolist]{acronym}
\usepackage{amsmath,amssymb,amsfonts}
\usepackage[hidelinks,bookmarks=false]{hyperref}
\usepackage{algorithmic}
\usepackage{graphicx}
\usepackage{textcomp}
\usepackage{xcolor}
\usepackage{orcidlink}
\usepackage{comment}
\usepackage{multirow}
\usepackage{float}

\def\BibTeX{{\rm B\kern-.05em{\sc i\kern-.025em b}\kern-.08em
    T\kern-.1667em\lower.7ex\hbox{E}\kern-.125emX}}
    
\begin{document}

\title{Energy-Efficiency Architectural Enhancements for Sensing-Enabled Mobile Networks}

\author{
\IEEEauthorblockN{
Filipe Conceicao\IEEEauthorrefmark{1}, Filipe B. Teixeira\IEEEauthorrefmark{2}, Luis M. Pessoa\IEEEauthorrefmark{2} and Sebastian Robitzsch\IEEEauthorrefmark{1}}
\IEEEauthorblockA{\IEEEauthorrefmark{1}InterDigital Europe Ltd, London, United Kingdom, Email: \{filipe.conceicao, sebastian.robitzsch\}@interdigital.com}
\IEEEauthorblockA{\IEEEauthorrefmark{2}INESC TEC and Faculdade de Engenharia, Universidade do Porto, Porto, Portugal,\\Email: \{filipe.b.teixeira, luis.m.pessoa\}@inesctec.pt}
}

\maketitle
\input{acronyms}

\begin{abstract}
Sensing will be a key technology in 6G networks, enabling a plethora of new sensing-enabled use cases. Some of the use cases relate to deployments over a wide physical area that needs to be sensed by multiple sensing sources at different locations. The efficient management of the sensing resources is pivotal for sustainable sensing-enabled mobile network designs. In this paper, we provide an example of such use case, and show the energy consumption due to sensing has potential to scale to prohibitive levels. We then propose architectural enhancements to solve this problem, and discuss energy saving and energy efficient strategies in sensing, that can only be properly quantified and applied with the proposed architectural enhancements. 
\end{abstract}


\section{Introduction}
Wireless sensing will be a pivotal technology enabler in 6G networks~\cite{IMT2030}, transforming a wide range of applications. By integrating advanced sensing capabilities with ultra-fast, low-latency 6G communications, wireless sensing will facilitate unprecedented levels of situational awareness and real-time responsiveness. It will enable more accurate environmental monitoring, enhanced user experience through immersive augmented and virtual reality, and smarter, more efficient urban management with intelligent infrastructures. Wireless sensors, embedded in various devices and infrastructure, will capture a wealth of data from environmental conditions to user interactions, enabling 6G networks to provide more personalized and context-aware services.

As sensing represents a new functionality in a wireless communication system, its energy consumption, which arises from data collection and processing, particularly in the context of \ac{IoT} devices, sensor networks, and mobile communications, must be a critical concern in modern sustainable wireless systems design. The process of sensing typically involves multiple stages, including identification of the sensing service and its location, sensing data acquisition via one or more sources, sensing data transmission and processing, each of which consumes varying amounts of energy. Data collection, often from distributed sensors, can be energy-intensive due to the need for continuous operation and high sampling rates. Wireless transmission further adds to energy demands, especially over long distances or in environments with poor signal quality, requiring more power for reliable communication. Once data is transmitted, processing is required~\cite{yao2023radar}, whether on-device or in an edge computing node or a central server, and that requires computational resources that also contribute to overall energy use. Otimizing these processes to reduce energy consumption is a big challenge, and typical strategies include data compression, efficient communication protocols, and low-power processing techniques to extend battery life and reduce the environmental impact of wireless systems \cite{yao2023radar}.

Smart transportation systems, including \ac{V2X} are a good example of sensing applications that need continuous environmental awareness via sensing monitoring, aggregating data across multiple sensors and locations. They enhance road safety, help optimizing traffic flows and reduce congestion. Additionally, they contribute to environmental sustainability by enabling more efficient driving patterns and reducing emissions. The integration of \ac{V2X} sensors in transportation infrastructure is crucial to the development of intelligent, connected, and safer road systems~\cite{cress2023intelligent} and multiple sources of sensors may be available, including cameras, \ac{LiDAR}, radar, RF sensing, and others~\cite{teixeira2024converge,CONVERGE-D1.1}. Sensor fusion techniques are required to create a more accurate, reliable, and comprehensive understanding of the environment than what could be achieved with any single sensor alone. However, the more sensing sources used to achieve the sensing goal, the higher the energy consumption, system complexity, processing capability requirements, and overall system usage is required. 

In this paper, we address the network sustainability aspects of sensing, by addressing its energy efficiency. We argue that different \ac{V2X} use cases present different requirements based on circumstances, location or the goal of sensing and show that system complexity and energy consumption may increase significantly to achieve sensing goals, and that this is a challenge that requires attention and further addressing. We then propose architectural enhancements to solve this problem, and discuss energy saving and energy efficient strategies.

\section{Related Work}
Based on the Paris Agreement from 2018, the mobile sector set the goal to reach net zero by 2050~\cite{gsma-netzero}. While 5G has significantly improved the energy efficiency in mobile networks over its predecessor, 6G must play a pivotal role in achieving the net zero target. And the challenge at hand is not just further improving on the operations towards more energy efficient networks, the challenge is also to further support the every growing demand for more data with an additional ``20 percent through 2029''~\cite{ericsson-mobiledataforecast}.

Sensing as an underlying functionality in 6G networks will increase energy consumption, especially in cases of wide area sensing deployments. The energy consumption will be due to wireless transmissions related to sensing, processing that is required on sensing related data, its storage, and the transmission of processed sensing data. Some work on energy efficiency can be found in~\cite{28310}, where concepts, use cases, requirements and solutions for the energy efficiency assessment and optimization for energy saving of 5G networks are presented. To the best of our knowledge, this is the best reference for architectural enhancements for energy efficiency in 6G networks, and little work as been presented addressing the problem of energy efficiency in sensing in wide area, real world scenarios.

\section{Sensing Data Sources in \ac{V2X}}
A large set of sensing data sources can be used in \ac{V2X} use cases, both to detect and interpret various physical phenomena and to derive environmental awareness. In this section, we discuss the strengths and drawbacks of three major sources: video sensing, \ac{RF} sensing and \ac{LiDAR}.

\subsection{Video-Based Sensing}
Video data streams for sensing involve the continuous capture and transmission of visual information through cameras. These are sequential frames of images that, when processed in real-time, provide a dynamic representation of the environment. In the context of autonomous vehicles, and using advanced processing techniques, including machine learning and computer vision, meaningful data can be extracted, such as vehicle, pedestrian or pattern detection, contributing to make informed decisions based on the computed results.

In the context of \ac{V2X} communication, the usage of video data streams being captured by a \ac{BS} and vehicles for sensing, depends on roles, capabilities, and integration of those nodes within the network. A \ac{BS}, typically fixed at a strategic location, captures a broader and more stable view of the environment, providing a macro-level perspective essential for traffic management, infrastructure monitoring, and emergency response coordination. Conversely, a vehicle captures video streams from a mobile, dynamic vantage point, offering a micro-level perspective crucial for real-time decision-making, obstacle detection, and situational awareness. While \ac{BS}s would aggregate and relay information to a central system for comprehensive analysis, vehicles can utilize on-board processing for immediate actions and point-to-point communication with other vehicles and infrastructure. These are advantages to be exploited, based on traffic context. The drawbacks include lower range, velocity and angle measurements capabilities compared to radar systems \cite{yao2023radar}, low latency, high reliability and high frame rate requirements for video transmissions which, in turn, consumes high amounts of bandwidth and energy, as well as the high computational cost associated with processing algorithms, for which processing capabilities need to be considered. 

\subsection{\acl{RF} Sensing}
\ac{RF} sensing involves the transmission and reception of \ac{RF} signals to detect and interpret various physical phenomena and to derive environmental awareness. By analyzing the reflected or scattered RF signals, it is possible to infer movement, detect objects and their dimensions, the object's positioning, and other characteristics. In \ac{V2X} systems, \ac{RF} can be used to accurately determine both the position and velocity of e.g., a vehicle. \ac{RF} data can be captured in mono-static, bi-static or multi-static modes, with either an access node, a vehicle or an active transceiver place along the road as the active transmitters and receivers~\cite{cress2023intelligent}. 

The drawbacks of RF sensing include high bandwidth usage requirements, object resolution problems, and susceptibility to noise and interference. The required bandwidth of the reference signal is typically proportional to the maximum Doppler frequency shift, which can create a significantly high bandwidth expenditure for higher speed vehicles, that then in turn cannot be used for communications with decoupled sensing and communications waveforms. Additionally, wider bandwidth reference signals require more complex signal processing techniques, which may not be available onboard.

\subsection{\acl{LiDAR} Sensing}
\ac{LiDAR} systems emit laser pulses and measure the time it takes for each pulse to return after hitting an object. High-resolution \ac{LiDAR} systems generate millions of data points per second, which equals to several GBit/s of raw data that needs to be processed for a real-time 3D point cloud computing of the scanned environment~\cite{maksymova2018review}. 

Despite the high resolution that can be achieved with \ac{LiDAR} systems, this data generation rate can cause significant challenges in real-time applications like \ac{V2X} where data must be processed instantly and real-time processing of \ac{LiDAR} data is essential for applications such as autonomous driving. This can lead to latency issues, affecting the system’s responsiveness, despite the various compression methods available~\cite{javed2020quick}. Transmission of these large volumes of data would require significant bandwidth usage. Additional drawbacks include lower range, velocity and angle measurement capabilities compared to radar systems~\cite{yao2023radar}, as well as performance degradation due to weather, in particular due to conditions such as bright sun, fog, rain, dirt and water spray.

\section{Sensing Energy Consumption}
Several advanced use cases for \ac{V2X} communication have been identified in the context of 6G networks, namely from important sources such as \ac{5G-AA}, \ac{3GPP}, \ac{5G-PPP} and European Union funded project efforts such as the CONVERGE and Hexa-X projects~\cite{3GPP_2024,CONVERGE-D1.1,Hexa-X_2021,5G-PPP_2019}. The \ac{3GPP} requirements for a \ac{V2X} collision avoidance between two vehicles use case are given as an example in Table~\ref{tablereq}~\cite{3GPP_2024}. 

\begin{table}[!ht]
\caption{\ac{V2X} use case requirements \cite{3GPP_2024}.}
\label{tablereq}
\begin{tabular}{|ll|l|l|}
\hline
\multicolumn{2}{|l|}{Sensing service area}                                                       & Outdoor          & Indoor           \\ \hline
\multicolumn{1}{|l|}{\multirow{2}{*}{Accuracy for position (m)}}            & Horizontal         & $\leq{1.3}$  & $\leq{2.6}$  \\ \cline{2-4} 
\multicolumn{1}{|l|}{}                                                      & Vertical           & $\leq{0.5}$  & $\leq{0.5}$  \\ \hline
\multicolumn{1}{|l|}{\multirow{2}{*}{Accuracy for velocity (m/s)}}          & Horizontal         & $\leq{0.12}$ & $\leq{0.12}$ \\ \cline{2-4} 
\multicolumn{1}{|l|}{}                                                      & Vertical           & N/A              & N/A              \\ \hline
\multicolumn{1}{|l|}{\multirow{2}{*}{Sensing resolution}}                   & Range (m)             & 0.4              & 0.4              \\ \cline{2-4} 
\multicolumn{1}{|l|}{}                                                      & Velocity (m/s)          & $\leq{0.6}$  & $\leq{0.6}$  \\ \hline
\multicolumn{2}{|l|}{\begin{tabular}[c]{@{}l@{}}Max Sensing Service latency (ms)\end{tabular}} & 50               & 20               \\ \hline
\end{tabular}
\end{table}

\begin{figure*}[!ht]
    \centering
\includegraphics[width=0.65\textwidth]{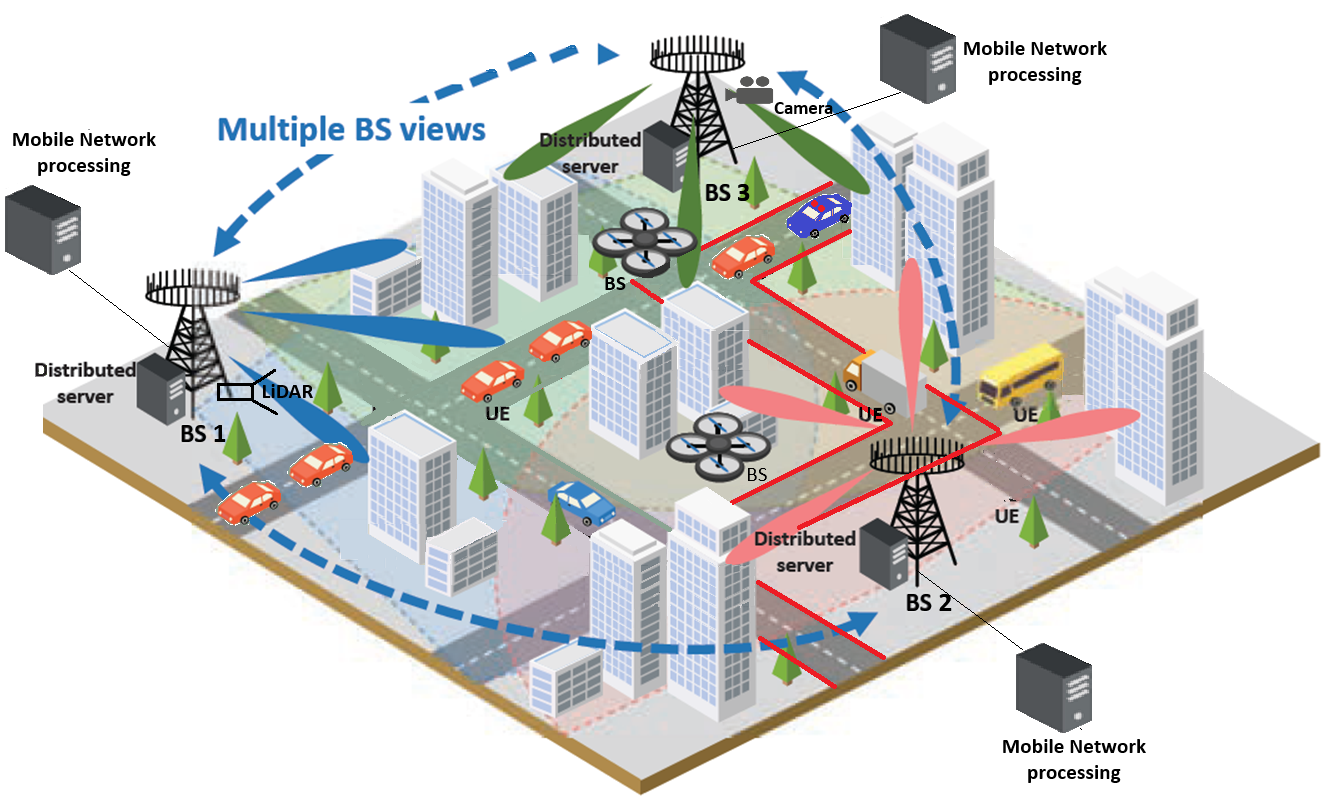}
    \caption{An emergency vehicle's route along a busy road, where the emergency vehicle is directed towards a less busy road.}
    \label{fig:v2x}
\end{figure*}

In the real world though, the scenarios can be substantially more complex. In a busy city, there are many daily occurrences of emergency response teams. Emergency vehicles and responding workers operate under strict time constrains, and it is of utmost importance to reach their destination as quickly as possible without any hurdles. 
In peak traffic hours, traffic bottlenecks to the emergency operation, increase delays and aggravate the emergency that triggered the emergency vehicle in the first place. Managing traffic of connected vehicles in such a scenario would be extremely beneficial in social terms, and would represent a major advancement in intelligent \ac{V2X} systems. Upon the detection of an emergency, the emergency vehicles' route can be calculated, and traffic can be moved ahead of its presence to clear the road for the fastest vehicle reachability possible, while minimizing the impact on the other vehicles. Fig.~\ref{fig:v2x} illustrates an example of this scenario. 

The figure also depicts how sensing could be applied to realise such a traffic management system. There are several \ac{BS} options: \ac{BS}1 is equipped with \ac{LiDAR} equipment, \ac{BS}2 with \ac{RF} capabilities only, and \ac{BS}3 with a camera. Vehicles on the road may have a mobile network compatible modem, and if so, connected. The vehicles are further equipped with \ac{LiDAR}, cameras and \ac{RF} capable equipment, and any generated data streams can be transmitted to a \ac{BS}. The \ac{BS}s have access to both a local edge compute nodes (distributed servers), and to processing resources (Mobile Network processing) beyond the \ac{RAN}, via backhauling. The emergency vehicle enters the area depicted in the figure, and faces a traffic jam at the main street. A traffic management application provides, in real-time, the vehicle with the fastest route to reach the emergency scene, depicted with red lines along the road, offloading remaining traffic to secondary routes if needed. 

The continuous usage of sensing functionalities in such a scenario across e.g., the roads of an entire city would result in a prohibitive energy consumption due to sensing. For example, it would be energy inefficient to continuously monitor an empty road that is behind the emergency vehicle's route, as opposed to one where there is the need to identify the number of vehicles, their types, speeds, positions on the road, as well as obstacles, pedestrians or animals for optimal traffic management or collision avoidance goals. In Fig.~\ref{fig:v2x} for example, the video resources from the camera in \ac{BS}3 might be sufficient to conclude the main road is jammed, allowing for activation of \ac{RF} sensing resources on \ac{BS}2, that can sense an empty road that is suitable for the emergency vehicle. \ac{BS}1 resources can be spared, both for \ac{RF} and \ac{LiDAR} capabilities. 

\subsection{Energy Efficiency Metrics}
Optimized management of sensing recources would increase energy efficiency. For an energy-efficient network design, \ac{EE} metrics are of primary importance because they form a basis for making optimized decisions across the different protocol layers.

\ac{EE} \ac{KPI}s typically measure a useful output of a functionality over the required energy expenditure to achieve it. Considering only the typical \ac{V2X} use cases have position and velocity requirements to be met in sensing, an \ac{EE} \ac{KPI} for this use case can be defined as Eq.~\eqref{EEsensingeq}.

\begin{equation}
EE=\sum_{i=0}^{Vehicles}\frac{a_{Pos}(n)+a_{Vel}(n)}{EC_{Sensing}(n)}\label{EEsensingeq}
\end{equation}

where $a_{Pos}$ and $a_{Vel}$ are the accuracy of position and velocity measurements for a vehicle (e.g., measured in \%), $n$ is the time step used, $EC_{Sensing}$ is the energy required to the useful output, and $i$ is the vehicle's index, and where detection and tracking of pedestrians, animals or objects on the road is not considered, for simplicity purposes. The time dependency is of high importance in such use cases due to the potential complexity of the physical scenario and the movement of vehicles along the planned route. The acquired accuracies via sensing will change over time, possibly with $n$ being the same as the sensing refresh rate requirement of a \ac{V2X} use case. 

The sensing functionality is then required to manage and maintain high levels for a metric like~\eqref{EEsensingeq} with a high confidence level (\(\geq 95\%\)) for position and velocity, and with low missed detection (\(\leq 10\%\)) and false alarm rates (\(\leq 1\%\)) \cite{3GPP_2024}. 

\subsection{Processing Energy Cost}
In Eq.~\eqref{ECsensingeq}, we further divide the Energy Consumption in sensing into the energy spent on sensing transmissions ($P_{Tx}$), where the consumption increases proportionally to the volume of data to be transmitted, and the Energy Consumption of network nodes that are used for processing of information ($P_{P}$). 

\begin{equation}
EC_{Sensing}(n)=P_{Tx}(n)+P_{P}(n)
\label{ECsensingeq}
\end{equation}

It is particularly important to note the cost of processing of, e.g., point cloud data, due to the need for GPU processing, that can be virtualised essentially in any hardware along the data stream's path. This also includes the possibility of slicing over a sensing operation for more efficient energy management in the end-to-end architecture. 

The task of reporting energy consumption metrics for sensing activities in a mobile network requires special technical considerations, which are discussed in herein. First, the consumed energy can be categorised into:
\begin{itemize}
    \item RF frontend: RF transceiver chains for performing sensing and exchanging sensing data wirelessly;
    \item Data transfer: fixed/cable transceiver chains for exchanging sensing data between compute nodes;
    \item Data process: pure softwarised network components to process sensing data.\label{enum:energy-cats-3}
\end{itemize}

For all three categories, the task of measuring precisely the energy consumed for a particular sensing task requires various standards to specify such capabilities, i.e., 3GPP for mobile networks, IEEE for Wi-Fi and Ethernet-based networks, and MPEG for video-related sensing activities. Furthermore, as described in Section~\ref{sec:Architecture}, 6G may see the adoption (and therefore extension) of \ac{SBA} principles in the \ac{RAN}, enabling a more unified and flexible mobile network. The core proposition of \ac{SBA} is the ability to scale the execution of software purely based on demand, assuming there is no compute, storage or networking limitations. Utilising cloud-native software engineering approaches where an \ac{SDO} could impose that sensing (or even individual sensing activities) are executed by dedicated software components, the measurement of the consumed energy for sensing activities becomes a rather straightforward task. Frameworks such as Scaphandre~\cite{scaphandre} allow to obtain the consumed energy per software process directly from the \acp{CPU}. Even hypervisors such as Linux QEMU are supported; however, only the power consumption for an entire \ac{VM} can be measured and not for individual processes inside the \ac{VM}. Hence, with tools such as Scaphandre and architectural principles such as \ac{SBA}, \acp{SDO} working on architectural specifications, e.g., 3GPP, play a vital role in enabling fine-grained energy consumption metrics for specific services. Furthermore, the \ac{NGMN} forum published a whitepaper on metering virtualised \ac{RAN} infrastructures \cite{MGMN-meteringNetworks}, elaborating further on the outlined challenge above, in particular on metering approaches in mobile networks including data centres in which softwarised components are deployed.

\section{Extension of 5G System Architecture for Energy Monitoring}\label{sec:Architecture}
In the \ac{V2X} context presented in this paper, it is argued that some real world examples present a high degree of complexity to achieve the sensing goal, especially under the presented constraints and additionally, in the presence of communications. A careful selection of the resources used, both for sensing transmission locations and its processing resources, needs to be ensured. As pointed out, sensing via any source can be very bandwidth demanding and generate huge volumes of data to be processed for meaningful results. With unlimited resources, all sensing stream types from all possible locations could be aggregated and processed, contributing to the highest possible position and velocity accuracy. However, besides the resource limitations, this would incur in a significant high energy cost. 

\begin{figure}[ht!]
  \centering
  \includegraphics[width=0.68\linewidth]{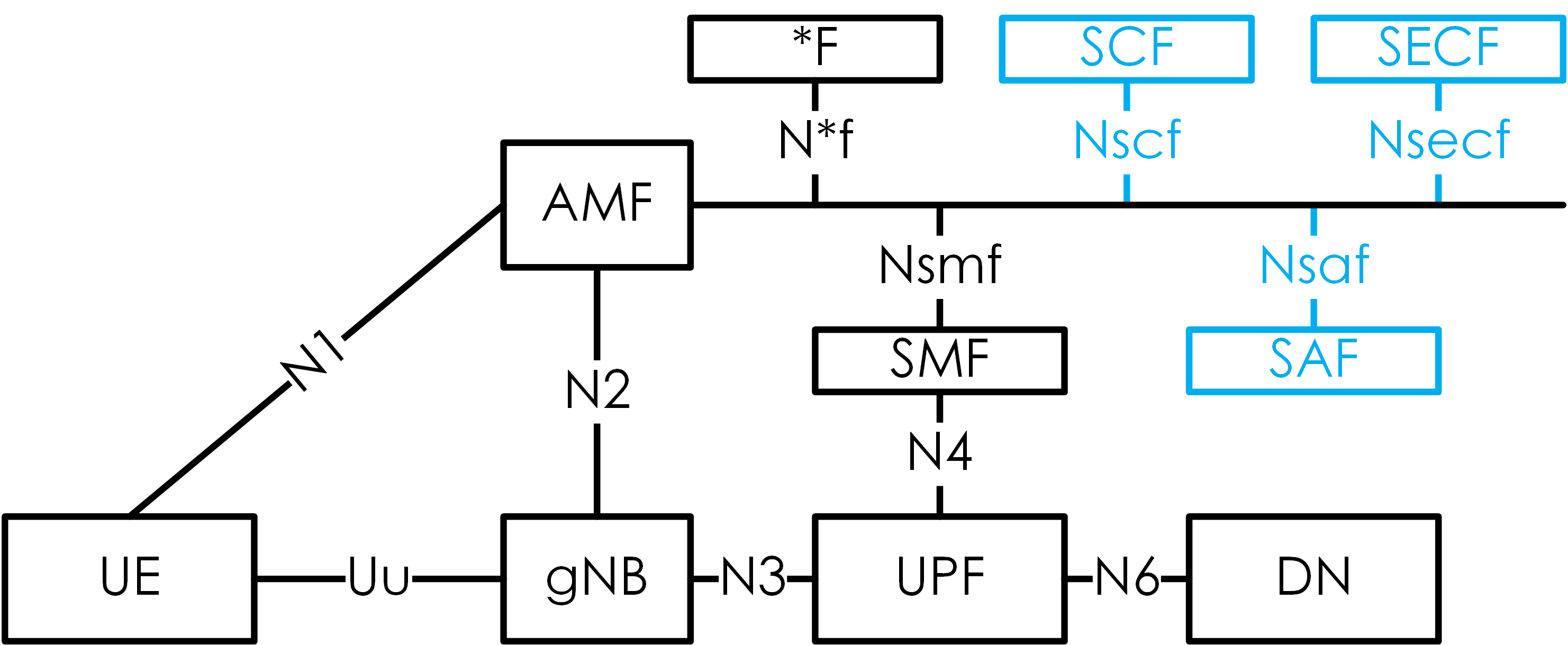}
  \caption{Energy Efficiency 5G architecture extension for 5G sensing scenarios.}
  \label{fig:Extended5GArch}
\end{figure}

\figurename~\ref{fig:Extended5GArch} illustrates an extended 5G system architecture enabling sensing in mobile networks for a range of sensor sources. The colored \acp{NF}, \ac{SCF}, \ac{SAF} and \ac{SECF}, form the proposed extension and are described in further detailed below. The extension of the system leverages \ac{SBA} principles where \ac{CN} functionality is separated into dedicated \ac{NF} and their interfaces unified in protocol usage (i.e., HTTP with JSON payload) and standardised in 3GPP. This enables operators to only purchase and deploy \ac{CN} functionality which is required in their network, e.g., sensing, and also enables multi-vendor deployments where \ac{NF} 1 is developed by Vendor A and \ac{NF} 2 by Vendor B. Also, \ac{SBA} principles enable unified monitoring of all \ac{NF} and to scale them on-demand using cloud-native software engineering approaches.

The newly proposed 5G \ac{CN} extensions are:
\begin{itemize}
    \item \ac{SCF}: coordinates sensing modes, sensing sources, and the execution of sensing tasks. This also includes the coordination of time dependencies of analytics related to energy efficiency;
    \item \ac{SAF}: performs analytics over sensing data and generates sensing results based on the collected sensing data. Furthermore, the \ac{SAF} performs fusion of sensing data from multiple sources. The results are then exposed to application servers in the \ac{DN} via the \ac{NEF} and \ac{AF};
    \item \ac{SECF}: collects, assesses and stores energy consumption data from all \ac{5GS} entities involved in sensing (including long term storage) and provides sensing results for the \ac{SCF} to improve on the energy consumption of the system.
\end{itemize}

Furthermore, it is expected that the \ac{PCF} holds energy efficiency policies to impose system behaviour such as performance versus energy consumption-aware. These policies are set by operators to fine the energy consumption of the system or requested by applications via their trusted \ac{AF}.

\section{Energy-Efficiency Sensing Strategies}
In this section, we discuss some energy efficiency strategies that could be put in place to mitigate energy expenditure from sensing, and we argue that they can only be evaluated and enforced with an energy efficiency monitoring functionality, capable of monitoring both sensing requirements fulfilment and energy consumption due to sensing. This entity needs to be able to combine situational awareness, energy consumption, sensing transmissions and processing resource usage assessment.

\subsection{Situational Awareness}
In the presented use case, the energy demand for sensing is significantly lower when the sensing area is smaller. This is however dependant on the road and route used, and its current load. For example, with a camera mounted on a \ac{BS} with a clear view of an empty road and its surroundings for a few hundreds of meters, the \ac{SAF} could conclude shorter range sensing around the emergency vehicle is sufficient for navigation. If the video stream covers a length of road served by two different access nodes, \ac{RF} sensing resources from the furthest node may not be required, and same applies to e.g., video or \ac{LiDAR} streaming from the vehicle. Equally, in a confined environment, like a parking space or a specific road intersection, the sensors need to cover a limited range, reducing the power needed for signal transmission and data processing. Additionally, fewer data points are generated in smaller areas, which means that the sensing computation and transmission overhead is also reduced. This contrasts with larger areas, such as an entire highway network or city, where the sensors operate over greater distances and handle more complex and voluminous data, leading to higher energy consumption. But if the route is empty and road conditions are optimal, this may not be required. Optimizing sensing tasks in \ac{V2X} systems by focusing on smaller, relevant areas for sensing can therefore contribute to more energy-efficient operations, extending the battery life of devices and reducing the overall power requirements of the system.

\subsection{Sensing Mode Optimisation}
Situational awareness also plays a role in the selection of sensing modes: mono-static, bi-static, or multi-static. The mode(s) selected to perform sensing can be influenced by the available video and/or \ac{LiDAR} sources, or work in isolation for \ac{RF} sensing operations. Each mode have distinct characteristics in terms of the sensing results they can achieve. Utilizing all modes for the presented \ac{V2X} use case would be counter-productive as the different modes cannot be used simultanously without negative consequences such as interference. For instance, mono-static sensing resource usage pertains to one device only, but it has the drawbacks of non-negligible interference in the transmission/reception chain, and a single reference point for angle of arrival estimation. The opposite is observed in bi-static mode, where interference effects are smaller but since there are more geographical reference points for angle estimation, the number of parameters to be estimated is higher~\cite{liu2022survey}, thus the overall system overhead for coordination is higher. These drawbacks scale upwards in multi-static, where the more nodes are involved, leading to higher the overhead and higher energy consumption. Assuming interference is negligible, using all sensing modes simultaneously (by different nodes) would yield the best results, but it would also lead to impractical energy consumption.

To balance energy use and sensing accuracy, mode selection can be dynamically adjusted based on the specific requirements of the situation. For instance, mono-static sensing might be preferred in low to no traffic environments, where either basic awareness suffices, or there is another sensing source supporting its usage. In contrast, multi-static sensing could be reserved for high-priority areas or critical situations, such as intersections with heavy traffic or areas prone to accidents, where maximum accuracy is necessary despite the possibly higher energy cost. Bi-static sensing could be deployed in scenarios requiring moderate accuracy, such as monitoring a specific road segment, where \ac{RF} sensing can be complimentary to in-vehicle mounted sensors.

\subsection{Time Dependencies}
The requirements described in Table~\ref{tablereq} are an example for a \ac{V2X} use case. There are, however, a myriad of other use cases where the sensing service latency requirements ranges between 10\,ms and 10 minutes~\cite{5G-AA_2020, 5G-PPP_2019,Hexa-X_2021}. This creates a broad range of time limits for sensing transmissions and processing to take place, and consequently, different possibilities for an efficient resource usage selection process, namely in their location. Use cases with lower latency requirements will have to consider processing of sensing generated data closer to the device that will take an action (e.g., nearby Edge or Fog compute nodes), while use cases with looser latency requirements can afford lower computational power, as there is more time for processing (e.g., in a more centrally-located instance of an \ac{SAF}). 

The location of compute resources to process sensing data will influence the energy consumption, and therefore its efficiency, due to two main reasons: 1) it will impact transmissions of sensing data across the system, and 2), it will dictate whether processing is done resorting to dedicated hardware, or virtualized software components running in a server. The location selection process is therefore not trivial because processing resorting to a central server would require more sensing data transmissions across the system, incurring into higher energy consumption. While computing closer to the device that will take action could potentially be a good strategy, drawbacks include the lower computational power, in addition to compute node load considerations, that can include sensing-related processing or other ongoing processes. A last time dependant requirement is the sensing result refresh rate. Similarly, this is another requirement with a broad range of values~\cite{5G-AA_2020, 5G-PPP_2019,Hexa-X_2021} for different use cases.

\section{Conclusions}
Sensing capabilitues are a key enabler to 6G networks. Achieving energy-efficient sensingis a complex challenge that demands advancements in hardware and algorithmic optimizations. 
We propose an Energy Efficiency 5G architecture extension for 5G sensing scenarios through the creation of  dedicated energy efficiency monitoring functionalities, presented as \ac{SECF}, and aided by \ac{SAF} and \ac{SCF}, that can oversee and help managing the interplay between sensing requirements and energy consumption by simultaneously monitoring the energy consumed during these operations. To be truly effective, this entity should integrate situational awareness, which allows it to adjust to changing environmental conditions, with a detailed assessment of the energy used in sensing transmissions and processing resources. By combining these elements, the system can make informed decisions on when and how to allocate resources to sensing tasks, thereby enforcing energy efficiency in real-time, as well as being able to quantify them. Without such a comprehensive monitoring and management framework, the goal of energy-efficient sensing in 6G networks would remain elusive, as there would be no mechanism to balance the trade-offs between sensing accuracy and energy usage dynamically~\cite{Heinzler_2019} \cite{aloufi2023object}. Future work includes the implementation of these functions to the CONVERGE architecture, to provide this research infrastrucuture energy efficiency capabilities.

\section*{Acknowledgment}
This work was supported by the CONVERGE project which has received funding under the European Union’s Horizon Europe research and innovation programme under Grant Agreement No 101094831.

\bibliographystyle{IEEEtran}
\bibliography{IEEEabrv,./references}

\end{document}

%% file: acronyms.tex
\acrodef{3GPP}{3rd Generation Partnership Project}
\acrodef{5G-AA}{5G-Automotive Association}
\acrodef{5G-PPP}{5G Infrastructure Public Private Partnership}
\acrodef{5GS}{5G System}
\acrodef{AF}{Application Function}
\acrodef{AR}{Augmented Reality}
\acrodef{CN}{Core Network}
\acrodef{CPU}{Central Processing Unit}
\acrodef{EE}{Energy Efficiency}
\acrodef{FR}{Frequency Range}
\acrodef{IoT}{Internet-of-Things}
\acrodef{KPI}{Key Performance Indicator}
\acrodef{LiDAR}{Light Detection and Ranging}
\acrodef{NEF}{Network Exposure Function}
\acrodef{NF}{Network Function}
\acrodef{NGMN}{Next Generation Mobile Networks}
\acrodef{PCF}{Policy Control Function}
\acrodef{RAN}{Radio Access Network}
\acrodef{RF}{Radio Frequency}
\acrodef{SAF}{Sensing Analytics Function}
\acrodef{SBA}{Service-Based Architecture}
\acrodef{SCF}{Sensing Coordination Function}
\acrodef{SECF}{Sensing Energy Consumption Function}
\acrodef{SDO}{Standardisation Development Organisation}
\acrodef{V2X}{Vehicle-to-Everything}
\acrodef{VAF}{Video Analytics Function}
\acrodef{VM}{Virtual Machine}
\acrodef{DN}{Data Network}
\acrodef{BS}{Base Station}

%% file: main.bbl
\begin{thebibliography}{10}
\providecommand{\url}[1]{#1}
\csname url@samestyle\endcsname
\providecommand{\newblock}{\relax}
\providecommand{\bibinfo}[2]{#2}
\providecommand{\BIBentrySTDinterwordspacing}{\spaceskip=0pt\relax}
\providecommand{\BIBentryALTinterwordstretchfactor}{4}
\providecommand{\BIBentryALTinterwordspacing}{\spaceskip=\fontdimen2\font plus
\BIBentryALTinterwordstretchfactor\fontdimen3\font minus \fontdimen4\font\relax}
\providecommand{\BIBforeignlanguage}[2]{{%
\expandafter\ifx\csname l@#1\endcsname\relax
\typeout{** WARNING: IEEEtran.bst: No hyphenation pattern has been}%
\typeout{** loaded for the language `#1'. Using the pattern for}%
\typeout{** the default language instead.}%
\else
\language=\csname l@#1\endcsname
\fi
#2}}
\providecommand{\BIBdecl}{\relax}
\BIBdecl

\bibitem{IMT2030}
\BIBentryALTinterwordspacing
I.~2030, ``Imt towards 2030 and beyond,'' 2023. [Online]. Available: \url{https://www.itu.int/en/ITU-R/study-groups/rsg5/rwp5d/imt-2030/Pages/default.aspx}
\BIBentrySTDinterwordspacing

\bibitem{yao2023radar}
S.~Yao, R.~Guan, Z.~Peng, C.~Xu, Y.~Shi, Y.~Yue, E.~G. Lim, H.~Seo, K.~L. Man, X.~Zhu \emph{et~al.}, ``Radar perception in autonomous driving: Exploring different data representations,'' \emph{arXiv preprint arXiv:2312.04861}, 2023.

\bibitem{cress2023intelligent}
C.~Cre{\ss}, Z.~Bing, and A.~C. Knoll, ``Intelligent transportation systems using roadside infrastructure: A literature survey,'' \emph{IEEE Transactions on Intelligent Transportation Systems}, 2023.

\bibitem{teixeira2024converge}
F.~B. Teixeira, M.~Ricardo, A.~Coelho, H.~P. Oliveira, P.~Viana, N.~Paulino, H.~Fontes, P.~Marques, R.~Campos, and L.~M. Pessoa, ``Converge: A vision-radio research infrastructure towards 6g and beyond,'' in \emph{2024 EuCNC \&6G Summit}.\hskip 1em plus 0.5em minus 0.4em\relax IEEE, 2024, pp. 1015--1020.

\bibitem{CONVERGE-D1.1}
\BIBentryALTinterwordspacing
``D1.1: Requirements and use cases,'' CONVERGE Project, 2023. [Online]. Available: \url{https://converge-project.eu/wp-content/uploads/2023/10/CONVERGE-D1.1-Requirements-and-use-cases-v1.0.pdf}
\BIBentrySTDinterwordspacing

\bibitem{gsma-netzero}
\BIBentryALTinterwordspacing
``Mobile net zero 2024 - state of the industry on climate action,'' GSMA, 2024. [Online]. Available: \url{https://www.gsma.com/solutions-and-impact/connectivity-for-good/external-affairs/climate-action/mobile-net-zero-2024/}
\BIBentrySTDinterwordspacing

\bibitem{ericsson-mobiledataforecast}
\BIBentryALTinterwordspacing
``Mobile data traffic outlook,'' Ericsson, 2024. [Online]. Available: \url{https://www.ericsson.com/en/reports-and-papers/mobility-report/dataforecasts/mobile-traffic-forecast}
\BIBentrySTDinterwordspacing

\bibitem{28310}
3GPP, ``3rd generation partnership project; technical specification group services and system aspects; management and orchestration; energy efficiency of 5g (release 18),'' 3GPP, Tech. Rep. TS 28.310 V18.5.0 (2024-06), 2024.

\bibitem{maksymova2018review}
I.~Maksymova, C.~Steger, and N.~Druml, ``Review of lidar sensor data acquisition and compression for automotive applications,'' in \emph{Proceedings}, vol.~2, no.~13.\hskip 1em plus 0.5em minus 0.4em\relax MDPI, 2018.

\bibitem{javed2020quick}
M.~Javed, M.~Meraz, and P.~Chakraborty, ``A quick review on recent trends in 3d point cloud data compression techniques and the challenges of direct processing in 3d compressed domain,'' \emph{arXiv preprint arXiv:2007.05038}, 2020.

\bibitem{3GPP_2024}
\BIBentryALTinterwordspacing
3GPP, ``Feasibility study on integrated sensing and communication (release 19),'' 2024. [Online]. Available: \url{https://www.3gpp.org/ftp/Specs/archive/22_series/22.837/22837-j40.zip}
\BIBentrySTDinterwordspacing

\bibitem{Hexa-X_2021}
\BIBentryALTinterwordspacing
Hexa-X, ``Deliverable d1.1 6g vision, use cases and key societal values,'' 2021. [Online]. Available: \url{https://hexa-x.eu/wp-content/uploads/2021/02/Hexa-X_D1.1.pdf}
\BIBentrySTDinterwordspacing

\bibitem{5G-PPP_2019}
\BIBentryALTinterwordspacing
5G-PPP, ``5g automotive vision,'' 2019. [Online]. Available: \url{https://5g-ppp.eu/wp-content/uploads/2014/02/5G-PPP-White-Paper-on-Automotive-Vertical-Sectors.pdf}
\BIBentrySTDinterwordspacing

\bibitem{scaphandre}
\BIBentryALTinterwordspacing
``Scaphandre - monitoring agent for energy consumption metrics,'' 2024. [Online]. Available: \url{https://github.com/hubblo-org/scaphandre}
\BIBentrySTDinterwordspacing

\bibitem{MGMN-meteringNetworks}
\BIBentryALTinterwordspacing
``Green future networks: Metering in virtualised ran infrastructure,'' NGMN, 2024. [Online]. Available: \url{https://www.ngmn.org/wp-content/uploads/GFN_Metering-Virt.-RAN-Infrastructure_V1.0.pdf}
\BIBentrySTDinterwordspacing

\bibitem{liu2022survey}
A.~Liu, Z.~Huang, M.~Li, Y.~Wan, W.~Li, T.~X. Han, C.~Liu, R.~Du, D.~K.~P. Tan, J.~Lu \emph{et~al.}, ``A survey on fundamental limits of integrated sensing and communication,'' \emph{IEEE Communications Surveys \& Tutorials}, vol.~24, no.~2, 2022.

\bibitem{5G-AA_2020}
\BIBentryALTinterwordspacing
5G-AA, ``C-v2x use cases volume ii: Examples and service level requirements,'' 2020. [Online]. Available: \url{https://5gaa.org/content/uploads/2020/10/5GAA_White-Paper_C-V2X-Use-Cases-Volume-II.pdf}
\BIBentrySTDinterwordspacing

\bibitem{Heinzler_2019}
\BIBentryALTinterwordspacing
R.~Heinzler, P.~Schindler, J.~Seekircher, W.~Ritter, and W.~Stork, ``Weather influence and classification with automotive lidar sensors,'' in \emph{2019 IEEE Intelligent Vehicles Symposium (IV)}.\hskip 1em plus 0.5em minus 0.4em\relax IEEE, June 2019. [Online]. Available: \url{http://dx.doi.org/10.1109/IVS.2019.8814205}
\BIBentrySTDinterwordspacing

\bibitem{aloufi2023object}
N.~Aloufi, A.~Alnori, V.~Thayananthan, and A.~Basuhail, ``Object detection performance evaluation for autonomous vehicles in sandy weather environments,'' \emph{Applied Sciences}, vol.~13, no.~18, 2023.

\end{thebibliography}
